\documentclass[10pt]{amsart}

\usepackage{amsmath}

\usepackage{latexsym}
\usepackage{amssymb,amsmath,amsfonts,enumerate}
\usepackage[matrix,arrow,arc]{xy}

\newtheorem{thm}{Theorem}[section]
\newtheorem{lemma}[thm]{Lemma}

\newtheorem{cor}[thm]{Corollary}
\newtheorem{defi}[thm]{Definition}
\newtheorem{ex}{Example}[section]

\newtheorem{remark}{Remark}[section]

\newcommand{\benu}{\begin{enumerate}}
\newcommand{\enu}{\end{enumerate}}

\newcommand{\beqna}{\begin{eqnarray}}
\newcommand{\eqna}{\end{eqnarray}}
\newcommand{\beqnast}{\begin{eqnarray*}}
\newcommand{\eqnast}{\end{eqnarray*}}
\newcommand{\beqn}{\begin{equation}}
\newcommand{\eqn}{\end{equation}}
\newcommand{\beqnst}{\begin{equation*}}
\newcommand{\eqnst}{\end{equation*}}

\newcommand{\N}{{\mathbb{N}}}

\newcommand{\bema}{\left ( \begin{array}}
\newcommand{\ema}{\end{array} \right )}

\newcommand{\E}{\mathcal{E}}

\newcommand{\orbit}{\mathcal{O}}

\newcommand{\F}{\mathbb{F}_q}

\newcommand{\Gl}[1]{\mathbf{Gl}_{#1}(\F)}
\newcommand{\T}[1]{\mathbf{T}_{#1}(\F)}
\newcommand{\M}{\mathbf{M}}

\begin{document}
\title[Generator Matrices of Codes in NRT Spaces]
{A standard form for generator matrices with respect to the Niederreiter-Rosenbloom-Tsfasman metric}
\author[M.M.S. Alves]{Marcelo Muniz S. Alves}\thanks{Partially supported by Funda\c{c}\~ao Arauc\'aria, 490/16032}
\address{Centro Polit\'ecnico, Departamento de Matem\'atica, Universidade Federal
do Paran\'a,  CP 019081, Jardim das Am\'ericas, Curitiba-PR,
81531-990, Brazil}
\email{marcelomsa@ufpr.br}

\maketitle

\begin{abstract}
In this note, we present an analogue for codes in vector spaces with a Rosenbloom-Tsfasman metric of the  well-known standard form of generator matrices for codes in spaces with the Hamming metric. 
\end{abstract}
\section{Introduction}
In this paper, $\F$ denotes the finite field of $q$ elements. 

A fundamental problem of coding theory is to estimate the highest minimum distance achievable by an $[n,k]_q$-code. Thinking in terms of 
parity-check matrices and Hamming metric, this is the same as estimating the highest $d \in \N$ such that there is an $(n-k) \times n$ matrix such that all its submatrices of $d-1$ columns have rank $d-1$. 
In \cite{N} and in some subsequent papers, H. Niederreiter studied a generalization of this problem to sets of vectors in $\F$-vector spaces. In \cite{brualdi}, Brualdi, Graves and Lawrence introduced the class of poset metrics and showed that Niederreiter's problem corresponds to the same fundamental problem of coding theory in terms of parity-check matrices, but with respect to a metric associated to disjoint unions of finite chains (i.e., finite totally ordered sets). 

A couple of years later, Rosenbloom and Tsfasman \cite{RT} introduced a metric over $n \times m$ matrices, known nowadays as the Rosenbloom-Tsfasman metric, which models a situation where errors occur in a specific burst pattern. It turns out that this metric turns is the particular case of the Niederreiter metric in which all chains have the same length $m$. Taking this into consideration, we will call this metric the Niederreiter-Rosenbloom-Tsfasman metric, or ``NRT metric'' for short. 

In this note, we present an analogue for NRT spaces (i.e., vector spaces with an NRT metric) of the  well-known standard form of generator matrices for codes in Hamming spaces, thus extending some results of \cite{chain}. In order to do this, we use the linear symmetries of NRT spaces.

\section{Niederreiter-Rosenblooom-Tsfasman Spaces}

Let $\M_{(m,n)}(\F)$ be the $\F$-vector space of $m \times n$ matrices with entries in  $\F$. The NRT weight on $\F^m$ is 
\beqnst
\omega (v) = \max \{i; v_i \neq 0\}
\eqnst
and the NRT weight on $\M_{(m,n)}(\F)$ is given by 
\beqnst
\omega ((v_{(i,j)})) = \sum_{i=1}^n \omega([v_{i,1} \ldots, v_{i,m}])
\eqnst
The NRT metric is the canonical metric associated to this weight, given by $d(u,v) = \omega(u-v)$. This metric is the \emph{poset metric} associated to the disjoint union of $n$ $m$-chains. We recall that if $P=\{x_1,\ldots,x_n\}$ is a finite poset, the \emph{ideal} generated by $X \subset P$ is $\langle X \rangle=\{x \in P; x \leq y \text{ for some } y \in X\}$. The $P$-weight on $\F^n$ is given by
\beqnst
\omega_P (v) = | \langle \operatorname{supp}(v) \rangle |  
\eqnst
where one identifies $x_i$ with its index $i$. The associated metric $d_P$ is the canonical one. 
 
Let $C_1, C_2, \ldots, C_n$ be distinct $m$-chains, and let $P = \bigcup_{j=1}^n C_j$. If we enumerate the elements of $C_j$ as $x_{(1,j)}, \ldots, x_{(m,j)}$, then it is easy to see that the $P$-metric on $\F^{mn} \simeq \M_{(m,n)}(\F)$ coincides with the NRT metric. Therefore, we will refer to the $P$-metric on $\F^{mn}$ as the NRT metric, and we will work with the space $\F^{mn}$ instead of the matrix space $\M_{(m,n)}(\F)$.

From \cite{groups,lee} one knows all the linear symmetries of $\F^{mn}$ with respect to  the NRT metric.
Let $\T{m}$ be the group of invertible upper-triangular $m \times m$ matrices, let $S_n$ be the symmetric group, and consider the direct product $(\T{m})^n = \T{m} \times \T{m} \times \cdots \times \T{m}$.  
Writing $\F^{mn}$ as $(\F^{m})^n $, where each $\F^m$ corresponds to one $m$-chain, $(\T{m})^n$ acts on $\F^{mn}$ by  
$$(T_1,T_2, \ldots, T_n) \cdot (v_1,v_2, \cdots, v_n) = (T_1v_1, \ldots, T_nv_n)$$
and $S_n$ acts by permuting components. It can be proved that both groups act faithfully by symmetries, and that  every linear symmetry of $\F^{mn}$ is a product of an element of $(\T{m})^n$ by a permutation. In fact, it can be proved that the group of linear symmetries is isomorphic to  the semidirect product $(\T{m})^n \rtimes S_n$, where  $S_n$ acts by permuting the components of $(\T{m})^n$. 

In a joint work with L. Panek and M.Firer  we introduced the concept of poset block spaces \cite{APF}, which include both spaces with an  error-block metric (defined in \cite{feng}) and spaces with a poset metric. Using this broader notion, in  \cite{chain} it is proved that every code in the generalized NRT space over 1 chain is equivalent to a trivial code. 

In the case of the usual NRT spaces, which are what concerns us in this work, this result says the following:

\bigskip

\begin{thm} \label{classif} \cite{chain}
If $C$ is an $[m,k]$-code in the NRT space $\F^{m}$ associated to an $m$-chain and $\beta=\{e_i\}_{i=1}^m$ is the canonical basis of $\F^m$, then $C$ is equivalent to the code defined by a generator matrix $G = [e_{i_1}; e_{i_2}; \ldots; e_{i_k}]$, where $1 \leq i_1 < i_2 < \ldots < i_k \leq n$ are the nonzero weights attained by codewords of $C$.
\end{thm}

\bigskip

%

In what follows, if $k \leq m$, we will identify $\T{k}$ with the subgroup 
\[
\left[
\begin{array}{c|c}
\T{k} & 0_{k, m-k} \\
\hline
0_{m-k,k} & I_{m-k}
\end{array}
\right]
\]
of $\T{m}$.

\bigskip

\begin{remark} Consider the action of $\T{m}$ on $\F^m$. \label{triangular} Analysing the corresponding matrices, it is easy to see that 
\begin{enumerate}[\rm(i)]
\item The stabilizer of $e_j$ is the subgroup of upper-triangular invertible matrices with $j$-th column equal to $e_j$.
\item The stabilizer of $e_j$ contains $\T{k}$ if $k < j$.
\item $\T{m}$ acts transitively on each $r$-sphere 
\[
S_r(0) = \{ \sum_{i=1}^r v_ie_i;   v_r \neq 0\}
\]
\item If $T \in \T{k}$, $1 \leq k <m$, and $v \in \F^m$ has its first $k$ coordinates equal to zero, then $Tv = v$. 
\end{enumerate}
\end{remark}

\bigskip

Surprisingly, $(\T{m})^n$ is transitive on the spheres of $\F^{mn}$  only in this case, where $n=1$ (see \cite{moscow} for  details). 
%
%
%
%

\section{Generator Matrices}
\begin{ex}  \label{example} Let $V=\mathbb{F}_2^8$ be given the poset metric associated to the disjoint union of two disjoint $4$-chains. Consider the binary $[8,4]$-code $C$ with generator matrix 
\beqnst
\left[
\begin{array}{cccc|cccc}
1 & 1 & 1 & 0 & 1 & 1 & 1& 1 \\
1 & 0 & 1 & 0 & 0 & 1 & 1& 0 \\
1 & 1 & 1 & 0 & 0 & 1 & 1& 1 \\
0 & 0 & 0 & 0 & 1 & 1 & 0& 0 \\
\end{array}
\right]
\eqnst
The submatrix which consists of the first four columns defines a code in a poset space over one chain. Using linear symmetries and elementary row operations, we can simplify this first submatrix, obtaining
\beqnst
\left[
\begin{array}{cccc|cccc}
0 & 1 & 0 & 0 & 1 & 1 & 1& 1 \\
0 & 0 & 1 & 0 & 0 & 1 & 1& 0 \\
0 & 0 & 0 & 0 & 1 & 0 & 0& 0 \\
0 & 0 & 0 & 0 & 1 & 1 & 0& 0 \\
\end{array}
\right]
\eqnst
which generates an equivalent code.
\end{ex}

\bigskip

The second submatrix (from the 5th to the 8th column) still has a lot of nonzero entries. The idea is to go on and simplify the second submatrix, without spoiling the first one. In what follows, we describe how to do this and we present a list of ``reduced forms '' for generator matrices. 

\bigskip 

Some words on notation:  given matrices $M_1 \in \M_{(r_1,t)}(\F),M_2 \in \M_{(r_2,t)}(\F),$ $ \ldots,
 M_n \in \M_{(r_n,t)}(\F)$ , the list $[M_1;M_2; \ldots:M_n]$, where these matrices appear separated by a semicolon, denotes the matrix whose first $r_1$ rows are formed by $M_1$, the following $r_2$ rows  are formed by the rows of $M_2$, and so on. In particular, $[v_1;v_2; \ldots;v_n]$ denotes the matrix whose $i$-th row is the vector $v_i$. 
 
 At the same time, we shall also need to describe a matrix in terms of its columns: If $A_1 \in \M_{(r,t_1)}(\F),M_2 \in \M_{(r,t_2)}(\F), \ldots,
 M_n \in \M_{(r,t_n)}(\F)$ , then $[M_1 M_2 \cdots M_n]$ or $[M_1 | M_2 | \cdots | M_n]$ denote (as usual) the matrix obtained by concatenating these matrices. 

We also have to define a notion of nondegeneracy for linear codes better suited to the NRT metric. 
Let $P$ be the disjoint union of $n$ $m$-chains, and consider the associated NRT space over $\F^{mn}$. We will say that a linear code in this space is \emph{nondegenerate} if the union of the supports of its codewords intersects each one of the $m$-chains. Note that this is weaker then the usual definition of nondegenerate code. 

Finally, when a group $G$ acts on a set $X$, the orbit of $p \in X$ will be denoted by $\orbit(p)$. 

\bigskip 

The action of $\T{m}$ on $\F^m $
induces  an action on elements of $\M_{(k,m)}(\F)$. Given $T \in \T{m}$ and  $M = [v_1; \ldots; v_k] \in \M_{(k,m)}(\F)$, this (left) action is given by 
\[
T \cdot M:= MT^t =  [Tv_1; \ldots; Tv_k].
\]
We will say that a $k\times m $ matrix $M$ is $\T{m}$-reduced 
if, modulo permutations of rows,  
\beqnst
M=[c_1 | c_2 | \cdots |c_m]
\eqnst
where each \emph{column} $c_j$ is either the zero vector or a vector of the form 
\[
c_j = (c_{(1,j)}, \ldots, c_{(\omega_j-1,j)}, 1, 0, \ldots , 0).
\]
with $\omega_j < \omega_{j'}$ whenever $j < j'$ and $c_j$ and $c_{j'}$ are nonzero. 

\bigskip 

Another group that appears naturally when we're trying to simplify generator matrices is the following: if $k = s_1 + s_2$, an 
elementary row operation is $(s_1,s_2)$-admissible if it is either
\begin{enumerate}
\item[(i)] a permutation of rows $v_i,v_j$, with both $i,j > s_1$, or  
\item[(ii)] a multiplication of a row $v_i$, $i>s_1$, by a nonzero scalar $\lambda \in \F$, or 
\item[(iii)] replacement of  $v_i$ by  $v_i + \lambda v_j$, with $j \geq s_1+1$. 
\end{enumerate} 
In other words, any elementary row operation can be performed on the submatrix $[v_{s_1+1}; \ldots; v_k]$ , and the only elementary operation allowed with rows $v_1, \ldots, v_{s_1}$ is multiplication by 
 \[
 I_s + \lambda E_{i,j} \text{ with } 1 \leq i \leq k \text{ and } s_1+1 \leq j \leq k. 
 \]
Writing each operation as multiplication by an elementary matrix on the left, and taking into account that the full set of elementary matrices generates the general linear group, we conclude that finite sequences of applications  of $(s_1,s_2)$-admissible row operations are in bijection with elements of  
 \beqnst
\E (s_1,s_2) = 
\left[
\begin{array}{c|c}
I_{s_1} & \M_{(s_1,s_2)}(\F) \\
\hline
0_{s_2,s_1} & \Gl{s_2}
\end{array}
\right]
\eqnst
which is a subgroup of $\Gl{k}$, acting by left multiplication on $\M_{(k,m)}(\F)$. Now, since this group acts by left multiplication and $\T{m}$ acts by right multiplication (after transposition), these actions commute with each other, and thus they define a left action of $G_{(s_1,s_2)} =  \T{m} \times \E (s_1,s_2)$ on $\M_{(k,m)}(\F)$ in the canonical way: given $(T,S) \in G_{(s_1,s_2)}$ and $M \in \M_{(k,m)}(\F)$, we define $(T,S) \cdot M = SMT^t$. 

\bigskip
 
\begin{lemma}\label{orbit} Let $M= [v_1;v_2; \ldots;v_k] \in \M_{(k,m)}(\F)$, let $k=s_1+s_2$ and consider the groups $\T{m}$ and $G(s_1,s_2)$ described above.  
\begin{enumerate}
\item[(i)] The orbit of $M$ under  $\T{m}$ contains a point $M_0$ which is $\T{m}$-reduced. 
\item[(ii)] If the last $s_2$ rows of $M$ are zero, then the orbit of $M$ under the action of $G_{(s_1,s_2)}$ has a $\T{m}$-reduced point
$M_0$ (with its last $s_2$ rows all zero);
\item[(iii)] If there is at least a nonzero row among the last $s_2$ rows of $M$, then its $G_{(s_1,s_2)}$-orbit contains a point 
$M_0=$
{\tiny
$ 
\left[
\begin{array}{cc}
A & B \\
J & 0 
\end{array}
\right]
$}, where
\begin{enumerate}
\item $J$ is a matrix whose nonzero rows are distinct canonical vectors, ordered from top to bottom by (increasing) NRT weight, and whose last column is nonzero
\item  
$A$ and $B$ are  $\T{m}$-reduced,  and all entries of $A$ above each nonzero entry of $J$
are zero. 
\end{enumerate}
\end{enumerate}
 \end{lemma} 

\bigskip

\begin{proof}
(i) By induction on $k$ (note that the case $k=1$ follows from Remark \ref{triangular}). 

Let $t \leq m$ be the maximum NRT weight of a row of the matrix $M= 
[u_1;u_2; \ldots; u_k] \in \M_{(k,m)}(\F)$. 
Without loss of generality, we may consider $t=m$ because if the $t$ is the maximum weight, then the action of  the triangular matrix 
\beqnst
T = 
\left[
\begin{array}{cc}
A_{t} & A_{t,m-t} \\
0_{m-t,t} & A_{m-t}
\end{array}
\right]
\in \T{m}
\eqnst
on $M$  coincides with the action of the matrix  
\beqnst
T' = 
\left[
\begin{array}{cc}
A_{t} & 0_{t,m-t} \\
0_{m-t,t} & I_{m-t}
\end{array}
\right]
 \in \T{t}
\eqnst

Modulo row permutations, we can assume that the last row $u_k$ has maximum weight. By Remark \ref{triangular}(iii), there is $T \in \T{m}$ such that $Tv_k = e_m$, and hence $T(M) =[v_1;v_2; \ldots; v_{k-1};e_m] \in \orbit(M)$. Let 
$j$ be the second least weight that appears in a row of  $T(M)$, i.e.,
{\small
\beqnst
T(M) = 
\left[
\begin{array}{ccc|cccc}
v_{1,1} &  \cdots  &v_{1,j} & 0 & \cdots &0& v_{1,m} \\ 
v_{2,1} &  \cdots  &v_{2,j} & 0 & \cdots &0& v_{2,m} \\ 
\vdots  &    &  &  &  & & \vdots  \\ 
v_{k-1,1} &  \cdots &v_{k-1,j} & 0 & \cdots &0& v_{k-1,m} \\ 
\hline 
0  &  \cdots & 0 & 0 & \cdots &0 & 1 \\ 
\end{array}
\right]
\eqnst}

\noindent where at least one of the entries of the $j$-the column is not zero. By the induction hypothesis, the upper left $(k-1)\times j$ submatrix is equivalent to a $\T{j}$-reduced matrix under the action of $\T{j}$. Since the action of elements of $\T{j}$ does not change the remaining columns (from $j+1$ to $m$), the result follows.

(ii)+(iii) If the rows from $s_1+1$ to $k$ are all zero, then the previous item implies that there is a point of type $[A;0]$ in the $G_{(s_1,s_2)}$-orbit of $M$. 

If there are nonzero rows with index strictly greater than $s_1$ then, by the classification for codes over one chain, it is possible to reduce the last $s_2$ rows to a sequence of distinct canonical vectors, which may be ordered by their indices,  and (possibly) some zero rows, thus obtaining a matrix $M'=${\tiny
$ 
\left[
\begin{array}{cc}
A' & B \\
J & 0 
\end{array}
\right]
$} where the nonzero rows of $J$ are distinct canonical vectors and its last column is nonzero. 

Finally, using $(s_1,s_2)$-admissible elementary row operations, we may kill off all entries of $A'$ above a nonzero entry of $J$, and we obtain a matrix 
$M_0=${\tiny
$ 
\left[
\begin{array}{cc}
A & B \\
J & 0 
\end{array}
\right]
$}
in the required form. 
\end{proof}
Since the action of $(\T{m})^n$ on $\F^{mn}$ corresponds, in terms of generator matrices, to the action of $(\T{m})^n$ on $\M_{(k,mn)}(\F)$ canonically induced by the previous action of $\T{m}$ on $\M_{(k,m)}(\F)$, an immediate consequence of the first item of the Lemma is

\bigskip

\begin{cor}Let $\F^{mn}$ be the NRT space over  the union of $n$ $m$-chains, and let $C $ be a linear $[mn,k]$-code. Then $C$ is equivalent to a code with a generator matrix $G=[G_1 | G_2 |\cdots |G_n]$, where each $G_i$ is a $\T{m}$-reduced $k \times m$ matrix.
\end{cor}

\bigskip

We can do something a little better than that:

\bigskip
\begin{defi}Let $\F^{mn}$ be the NRT space over  the union of $n$ $m$-chains, and let $C $ be a linear $[mn,k]$-code. We will say that a generator matrix $G$ for $C$ is in NRT-triangular  form if  $G=[G_1 | G_2 |\cdots |G_n]$, each $G_i$ a $k \times m$ matrix, where
\begin{enumerate}[(1)]
\item  $G$ is in ``block echelon form'', i.e., if the last $t$ rows of $G_i$ are zero, then the last $t$ rows of $G_1, \ldots, G_{i-1}$ are also zero; 
\item  The nonzero rows of $G_1$ are distinct canonical vectors, arranged in order of increasing NRT weight;
\item For each $i=2, \ldots, n$, 
\begin{enumerate}
\item[(i)] $G_i$ is $\T{m}$-reduced, or 
\item[(ii)]$ G_i=$
{\tiny
$ 
\left[
\begin{array}{cc}
A^i & B^i \\
J^i & 0 
\end{array}
\right]
$}, 
 where $A^i$ and $B^i$ are  $\T{m}$-reduced,  $J^i$ is a matrix whose nonzero rows are distinct canonical vectors (also arranged in order of increasing NRT weight) and whose last column is nonzero, and all entries of $A^i$ above each nonzero entry of $J^i$
are zero. 
\end{enumerate}
\end{enumerate}
\end{defi}

\bigskip 

\begin{thm}\label{generator}
Let $\F^{mn}$ be the NRT space over  the union of $n$ $m$-chains, and let $C $ be a linear $[mn,k]$-code. Then $C$ is equivalent to a code with has a generator matrix $G=[G_1 | G_2 |\cdots |G_n]$ in NRT-triangular form.
\end{thm}

\bigskip

\begin{proof}
Let $M$ be any generator matrix for $C$. This matrix can be written as $M=[M_1 | M_2 | \cdots | M_n]$, where each block $M_j$ contains the coordinates which correspond to the $j$-th $m$-chain. 

Permutations of chains induce permutations of these submatrices, so we may assume that $M_1$ is already a submatrix of minimum rank $r_1$ (among the $G_j$'s). By Theorem \ref{classif} for codes over one chain (or by Lemma \ref{orbit} with the action of $G_{(0,k)}$), $M_1$ can be reduced, via isometries and row operations, to a matrix which is a column permutation of a matrix of type
\beqnst
\left[
\begin{array}{cc}
I_{r_1} & 0_{r_1 \times (k - r_1)} \\
0_{(k - r_1) \times r_1} & 0_{k} 
\end{array}
\right]
\eqnst
and then $M$ can be replaced by a new matrix which begins with this submatrix and is generates a code equivalent to $C$. By simplicity, let us denote this new matrix also by $M$.

Write $M_j = [M_j^1; M_j^2]$, where $M_j^1$ consists of the first $r_1$ rows and $M_j^2$ of the following $k-r_1$ rows  of $M_j$. Permuting the submatrices if necessary, we may assume that $M_2^2$ has minimum rank among all $M_j^2$'s, $j \geq 2$ ($M_2^j$ might even be a null matrix). 

Note that $\T{m}$ acts on $M_2$ leaving all other submatrices $M_j$ fixed, and that every $(r_1,k-r_1)$-admissible elementary row operations on $M$ leaves $M_1$ fixed (but not the other $M_j$'s, though). Hence, Lemma \ref{orbit} says that $M_2$ is equivalent to a matrix of  type (3.i) or (3.ii). Proceeding in this manner, we obtain a matrix $G$ which generates a code equivalent to $C$ and in the form prescribed.
\end{proof}

\bigskip

\begin{ex}
In Example \ref{example} we arrived at the matrix
\beqnst
\left[
\begin{array}{cccc|cccc}
0 & 1 & 0 & 0 & 1 & 1 & 1& 1 \\
0 & 0 & 1 & 0 & 0 & 1 & 1& 0 \\
0 & 0 & 0 & 0 & 1 & 0 & 0& 0 \\
0 & 0 & 0 & 0 & 1 & 1 & 0& 0 \\
\end{array}
\right]
\eqnst
and, applying the method described above, we obtain the generator matrix
\beqnst
\left[
\begin{array}{cccc|cccc}
0 & 1 & 0 & 0 & 0 & 0 & 0& 1 \\
0 & 0 & 1 & 0 & 0 & 0 & 1& 0 \\
0 & 0 & 0 & 0 & 1 & 0 & 0& 0 \\
0 & 0 & 0 & 0 & 0& 1 & 0& 0 \\
\end{array}
\right]
\eqnst
which generates an equivalent code and has more zero entries than the original one. 
\end{ex}

Some interesting subcases follow from Theorem \ref{generator}. First of all, note that if the chains have length $m=1$ then this procedure leads to a generator matrix in standard form for the Hamming metric. If we work with two chains, we get 

\bigskip 

\begin{cor}[Codes over two chains] 
Let $\F^{m \times 2}$ be the NRT space over  the union of $2$ $m$-chains, and consider a linear code $C \subset \F^{m \times 2}$ of dimension $k$.  Then $C$ has a generator matrix $G$ of the form (i) $[A_1 | A_2]$ or (ii) 
{\tiny
$\left[
\begin{array}{cc}
A_1 & A_2 \\
0 & A_3
\end{array}
\right]$},
where
\begin{enumerate}[(1)]
\item the nonzero rows of $A_1$ are distinct canonical vectors, arranged in order of increasing NRT weight;
\item $A_2$ is $\T{m}$-reduced;
\item If $G$ is in the second form, then $A_3$ also consists of  distinct canonical vectors, arranged in order of increasing NRT weight, and all entries of $A_2$ above a nonzero entry of $A_3$ are zero. 
\end{enumerate}
\end{cor}

\bigskip 

Bidimensional codes also have a simple NRT-triangular form. 

\bigskip 

\begin{cor}[Bidimensional Codes] Let $\F^{m \times n}$ be the NRT space over  the union of $n$ $m$-chains, and let $C \subset \F^{m \times n}$ be a bidimensional code. Then $C$ has a generator matrix of the form 
\beqnst
\left[
\begin{array}{cccc}
G_1 & G_2 & \ldots & G_n
\end{array}
\right]
\eqnst
where each $G_j$ is a $2 \times n$ matrix of one of the following types:
null matrix, $ [e_i;0]$, $[0;e_i]$,  $[e_k;e_j]$ with $k \neq j$, or  $[e_k + \lambda e_j;e_j]$, where $1 \leq k < j$ and $\lambda \neq 0$. 
\end{cor}


 \makeatletter \renewcommand{\@biblabel}[1]{\hfill#1.}\makeatother

\end{document}